# Suppression of Chaos in Mutually Coupled Synchronized Generalized Lorenz Systems

V. RamiyaGowse[1], B. Palanivel[1], S.V.M.Satyanarayana[2], S.Sivaprakasam[2,*]

[1]*Department of Physics, Pondicherry Engineering College, Puducherry, India.*

[2]*Department of Physics, Pondicherry University, Puducherry, India.*

{ramiyaanand@yahoo.in, bpvel@pec.edu, svmsatya@gmail.com, siva@iitk.ac.in[*]}

**Abstract**

In this work, the dynamics of a system of mutually coupled Generalized Lorenz systems (GLS) is investigated. The state variables of two Lorenz oscillators are coupled mutually via non-linear controls and synchronization is achieved between the state variables. We find that by suitably controlling a parameter having a bearing on the coupling coefficient between the two Lorenz oscillators, the GLS, while preserving synchronization is rendered to a state wherein chaotic nature of state variables is suppressed and state variables exhibit oscillatory character. The suppression of chaos is verified by power spectra, permutation entropy and Lyapunov exponent calculations. When operated in chaotic domain, we show the possibility of transition from the state of synchronization to the state of anti-synchronization.

## 1. Introduction

A great deal of research had been carried out on chaos synchronization and control of chaos in various systems due to their applicability in multiple disciplines of research and applications [1-13]. The dynamical behavior of any chaotic system is known to be dependent sensitively on the initial values and the attractors and phase portraits are invariant when the control parameters are fixed. Later studies have revealed that certain nonlinear dynamical systems can be switched between chaotic to periodic attractors by certain control methods. In other words, the chaos could be controlled. Two broad categories of chaos control are feedback method and non feedback methods. Examples of feedback methods include the Ott-Grebogi-Yorke (OGY) method [14], occasional proportional feedback (OPF) method, time-delayed feedback control method (TDFC) etc. Examples of non-feedback methods include adaptive control, resonant parametric perturbation, weak perturbation entrainment and migration control, etc [15-19].

In this paper we present our investigations of synchronization properties of two mutually coupled Lorenz oscillators [20, 21], wherein the coupling is enabled via non-linear controls. In the past suppression of chaos has been reported by means of external driving [22-25]. We show in this work, that suppression of chaos is possible in a system of coupled Lorenz oscillators exhibiting synchronization. This is achieved by varying a control parameter having a bearing on coupling strength between the two oscillators. Power spectra, phase portraits, largest Lyapunov exponents and permutation entropy calculations confirm the suppression of chaos. In this work, we also show the co-existence of synchronization and anti-synchronization between state variables of both the oscillators for appropriate control parameter in the control functions enabling the mutual coupling between the two oscillators.

## 2. System Description

We consider two Lorenz oscillators and they are mutually coupled via non-linear controls [26]. Lorenz oscillator -1 (LO-1) is described by the system of following equations:

$$\dot{x}_1 = a(x_2 - x_1) + u_1$$
$$\dot{x}_2 = bx_1 + dx_2 - x_1 x_3 + u_2$$
$$\dot{x}_3 = x_1 x_2 + cx_3 + u_3 \qquad (1)$$

and the second Lorenz oscillator (LO-2) is described as follows :

$$\dot{y}_1 = a(y_2 - y_1) + u_4$$
$$\dot{y}_2 = by_1 + dy_2 - y_1 y_3 + u_5$$
$$\dot{y}_3 = y_1 y_2 + cy_3 + u_6 \qquad (2)$$

Where $x_i$ and $y_i$ (i = 1, 2, 3) are the state variables, $u_1, u_2, u_3$ are the control functions coupling the LO-1 with LO-2 while $u_4, u_5, u_6$ are the control functions coupling the LO-2 with LO-1.

The parameter values are assigned as,

$$a = 10 + \frac{25}{29}k; b = 28 - \frac{35}{29}k; c = -\frac{8}{3} - \frac{1}{87}k, d = k-1$$

The control functions are defined by the following expressions [26]:

$$u_1 = -(b + \sigma_3 x_3)E_2 + \sigma_2 x_2 E_3$$
$$u_2 = (\sigma_1\sigma_3 + \sigma_2)x_1 x_3 + (E_3 - a)E_1 - 2cE_2$$
$$u_3 = -(\sigma_1\sigma_2 + \sigma_3)x_1 x_2 - E_1 E_2$$
$$u_4 = -(b + \sigma_3 y_3)E_2 + \sigma_2 y_2 E_3$$
$$u_5 = (\sigma_1\sigma_3 + \sigma_2)y_1 y_3 + (E_3 - a)E_1 - 2cE_2$$
$$u_6 = -(\sigma_1\sigma_2 + \sigma_3)y_1 y_2 - E_1 E_2 \qquad (3)$$

Where, $(\sigma_1, \sigma_2, \sigma_3) = \sigma$ are the control parameters of the system, influencing the dynamics of coupling between the two oscillators and for which numerical values are assigned while carrying out the simulations. Here, $E_i$ (i=1,2,3) is the error function defined as

$$E_i = y_i + \sigma_i x_i; i = 1,2,3 \qquad (4)$$

Substituting eqn. (3) in (1), we get,

$$\dot{x}_1 = a(x_2 - x_1) - (b + \sigma_3 x_3)E_2 + \sigma_2 x_2 E_3$$
$$\dot{x}_2 = bx_1 + dx_2 - x_1 x_3 + (\sigma_1\sigma_3 + \sigma_2)x_1 x_3 + (E_3 - a)E_1 - 2cE_2$$
$$\dot{x}_3 = x_1 x_2 + cx_3 - (\sigma_1\sigma_2 + \sigma_3)x_1 x_2 - E_1 E_2 \qquad (5)$$

Substituting eqn. (3) in (2), we get,

$$\dot{y}_1 = a(y_2 - y_1) - (b + \sigma_3 y_3)E_2 + \sigma_2 y_2 E_3$$
$$\dot{y}_2 = by_1 + dy_2 - y_1 y_3 + (\sigma_1\sigma_3 + \sigma_2)y_1 y_3 + (E_3 - a)E_1 - 2cE_2$$
$$\dot{y}_3 = y_1 y_2 + cy_3 - (\sigma_1\sigma_2 + \sigma_3)y_1 y_2 - E_1 E_2 \qquad (6)$$

The error dynamical system is defined as

$$\dot{E}_i = \dot{y}_i + \sigma \dot{x}_i; i = 1,2,3 \qquad (7)$$

We take $(\sigma_1, \sigma_2, \sigma_3) = \sigma$,

As per (7), adding (5) and (6), we get,

$$\dot{E}_1 = -aE_1 + (a - b - \sigma b)E_2 - (1 - b\sigma)E_2 E_3$$

$$\dot{E}_2 = (b - a(1+\sigma))E_1 + (d - 2c(1+\sigma))E_2 + (1+\sigma)E_1 E_3 + (\sigma^2 + \sigma - 1)y_1 y_3 - \sigma x_1 x_3$$

$$\dot{E}_3 = cE_3 - (\sigma+1)E_1 E_2 - (\sigma^2 + \sigma - 1)y_1 y_2 - \sigma(\sigma^2 + \sigma - 1)x_1 x_2 \qquad (8)$$

In order to study the stability of $(E_1, E_2, E_3) = (0, 0, 0)$, the system of equations in (8) are linearized as follows,

$$\dot{E}_1 = -aE_1 + (a - (\sigma + 1)b)E_2$$

$$\dot{E}_2 = (b - (\sigma + 1)a)E_1 + (d - 2c(\sigma + 1))E_2$$

$$\dot{E}_3 = cE_3 \qquad (9)$$

We perform Lyapunov stability analysis and is as follows:

Consider the following Lyapunov function:

$$V = \frac{1}{2}\left(E_1^2 + E_2^2 + E_3^2\right) \qquad (10)$$

Differentiating V, with respect to time (t), we get,

$$\dot{V} = E_1\dot{E}_1 + E_2\dot{E}_2 + E_3\dot{E}_3 \qquad (11)$$

$$= E_1[-aE_1 + (a-(\sigma+1)b)E_2] + E_2[(b-(\sigma+1)a)E_1 + (d-2c(\sigma+1))E_2] + E_3[cE_3]$$
$$= -aE_1E_1 + (a-(\sigma+1)b)E_1E_2 + (b-(\sigma+1)a)E_2E_1 + (d-2c(\sigma+1))E_2E_2 + cE_3E_3$$

In matrix form, the above equation can be written as,

$$\dot{V} = (E_1 \quad E_2 \quad E_3)\begin{bmatrix} -a & a-(\sigma+1)b & 0 \\ b-(\sigma+1)a & d-2c(\sigma+1) & 0 \\ 0 & 0 & c \end{bmatrix}\begin{pmatrix} E_1 \\ E_2 \\ E_3 \end{pmatrix}$$

$$= -(E_1 \quad E_2 \quad E_3)Q(E_1 \quad E_2 \quad E_3)^T \qquad (12)$$

Where, $Q = \begin{bmatrix} a & (\sigma+1)b-a & 0 \\ (\sigma+1)a-b & 2c(\sigma+1)-d & 0 \\ 0 & 0 & -c \end{bmatrix}$

To ensure that the origin of the error system is asymptotically stable, we let the matrix Q be positive definite. This is the case if and only if the following three conditions hold:

$(i)\, a. > 0$

$(ii)\, \begin{vmatrix} a & (\sigma+1)b-a \\ (\sigma+1)a-b & 2c(\sigma+1)-d \end{vmatrix} > 0$

$(iii)\, \begin{vmatrix} a & (\sigma+1)b-a & 0 \\ (\sigma+1)a-b & 2c(\sigma+1)-d & 0 \\ 0 & 0 & -c \end{vmatrix} > 0 \qquad (13)$

All the three cases are verified for specific values of σ and hence for the system under study, the condition $\dot{V} > 0$ is satisfied, which implies that the error function is asymptotically stable at origin. In the next section we present our numerical findings. The fourth order Runge-Kutta algorithm is used to obtain solutions of equations (1) and (2). The parameters chosen are: step-size (h) =0.05, k=0.5, initial values of the variables are set as $(x_1, x_2, x_3)^T = (0.999, 0.899, 0.799)^T$ and $(y_1, y_2, y_3)^T = (1.0, 1.0, 1.0)^T$.

## 2. SUPPRESSION OF CHAOS

Lorenz oscillators LO-1 and LO-2 are coupled to each other as defined in equations (1) and (2). The parameters are chosen such that the system is in the chaotic regime. The strength of coupling between the two oscillators LO-1 and LO-2 are controlled by introducing a scale factor (sf) for the control functions $u_i$ (i = 1 to 6). The scale factors are designated as $sf_1$ and $sf_2$ scaling $u_i$ (i=1, 2, 3) and $u_i$ (i=4, 5, 6) respectively. The values of $sf_1$ and $sf_2$ are varied and its influence on the dynamics of the coupled system is studied while synchronization between the two oscillators is preserved. For simplicity, we have kept the scale factors $sf_1$ and $sf_2$ to be equal. For example, for a scale factor of 0.07, the Lorenz oscillators are found to exhibit the state of chaos synchronization as shown in Figure 1(a), in which we have shown the time evolution of the third state variable (either $x_3$ or $y_3$). We increased the scale factor in smaller steps and for example at a scale factor of 0.082, as shown in figure 1(b) has an initial signature of suppression of chaos though chaotic dynamics prevails as time evolves. As the scale factor is increased to 0.083, chaos is completely suppressed as shown in figure 1(c). As the scale factor is increased further to 0.14, the system exhibits damped relaxation oscillation, which is a signature of stabilisation of the system, in other words, the system has lost its chaotic behaviour.Thus it is argued that the oscillators which are intrinsically chaotic, when coupled appropriately mutually suppress chaos while preserving synchronization. This can also be seen as a method of controlling chaos of dynamical systems such as GLS by enabling mutual coupling between the individual oscillators using appropriate control function and scale factors. We further increased the scale factor and time evolution of LO-1 and /or LO-2 in figures 1(c-d) wherein chaos suppression is evident.

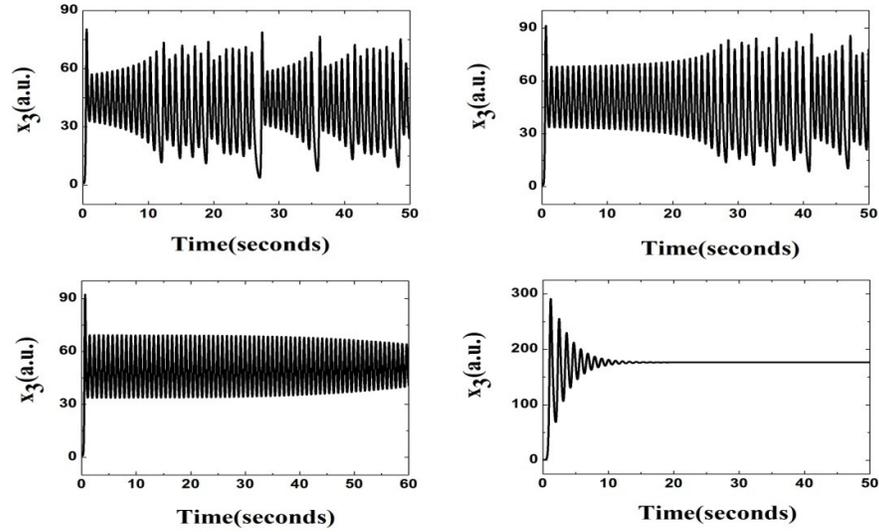

**Figure 1**: Time evolution of the state variable $x_3$, of either LO-1 or LO-2 (they are synchronized under mutual coupling) for different scale factors: viz., **(a)** scale factor =0.07, **(b)** scale factor = 0.082, **(c)** scale factor=0.083, **(d)** scale factor =0.14.

The corresponding phase portraits plotted between the state variable $x_3$ and its derivative are shown in figure 2.

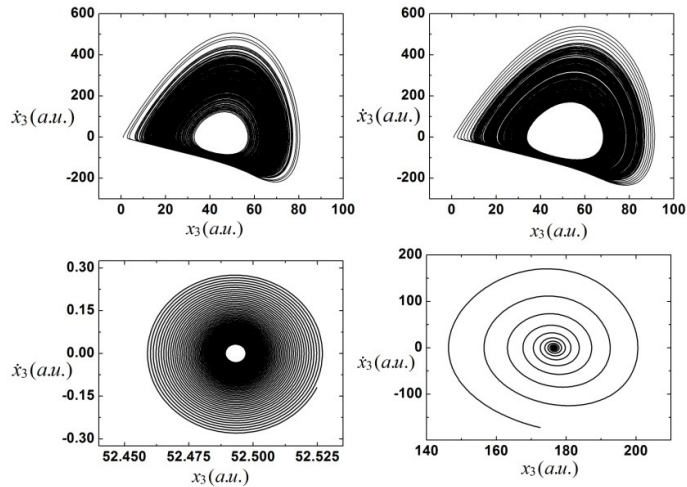

**Figure2**: Phase portraits of the state variable $x_3$, of either LO-1 or LO-2 (they are synchronized under mutual coupling) for different scale factors: viz., **(a)** scale factor =0.07, **(b)** scale factor = 0.082, **(c)** scale factor=0.083, **(d)**scale factor = 0.14.

While the portraits presented in 2(a) and 2(b) indicates the presence of chaotic nature, the portraits in 2(c) and 2(d) confirms the presence of undamped and damped oscillations respectively.

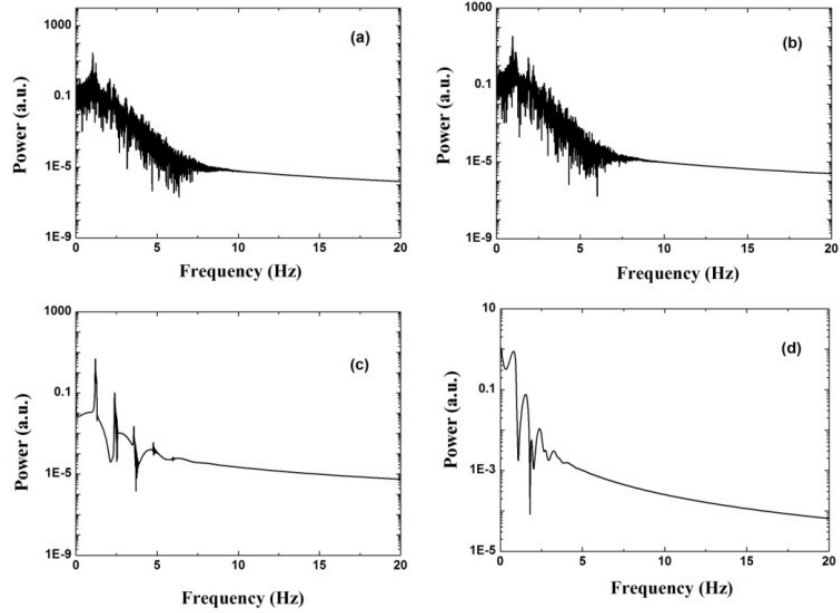

**Figure 3**: Power spectra of the state variable $x_3$, of either LO-1 or LO-2 (they are synchronized under mutual coupling) for different scale factors: viz., **(a)** scale factor =0.07, **(b)** scale factor = 0.082, **(c)** scale factor=0.083, **(d)** scale factor = 0.14.

Figure 3(a-d) shows the power spectra of time evolution of either of the oscillators (both are synchronized). Figures 3(a) and 3(b) exhibits multiple high frequency oscillations, evidently implying the existence of chaotic nature of the system. From figures 3(c) and 3(d) it is evident that chaotic nature is suppressed and periodicity is exhibited.

We calculated the largest Lyapunov exponent [27-29] so as to quantify the chaotic nature of oscillators. For this we consider two nearby trajectories obtained from the time series data, with their initial amplitudes ($x$) being close to each other at times say $t_i$ and $t_j$. Now, we consider the sequence of oscillator evolution of the state variable at times $t_i, t_{i+1}, t_{i+2}, \ldots$ and $t_j, t_{j+1}, t_{j+2}, \ldots$ and we find the divergence of these two sequences: divergence(div) = $|x_{i+d} - x_{j+d}|$; d = 0,1,2... If the system is chaotic, the plot of div vs. time will rise exponentially. For this, we plot ln(divergence) vs. time and apply a linear fit. The slope is an estimate for the Lyapunov exponent. Figure 4 (a) shows the typical plot of <ln(divergence)> versus time for the time series data at two different scale factors. The slope of linear fit provides the largest lyapunov exponent. The largest lyapunov exponent ($\lambda_{max}$) calculated for different values of scale factor are shown in figure 4(b). The maximum Lyapunov exponent is found to decrease as the scale factor is increased implying that the chaotic nature is being suppressed.

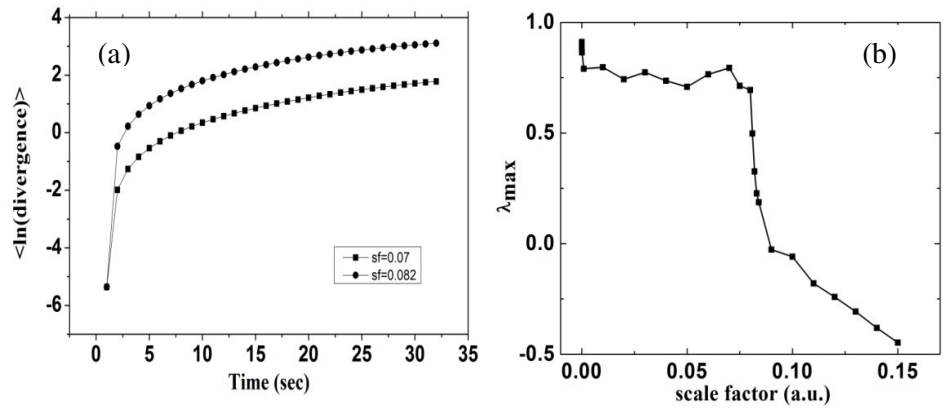

**Figure 4(a)**: <ln(divergence)> versus time for the mutually coupled GLS. **Figure 4(b)**: Largest lyapunov exponent ($\lambda_{max}$) calculated at different scale factor values.

Permutation entropy [PE] calculations [30-32] are used to quantify the degree of chaos. The normalized value (H) of Permutation entropy can vary from 0 to 1 ($0 \leq H \leq 1$), with H=0 corresponding to completely predictable dynamics and H=1 corresponding to completely unpredictable dynamics. At each of the scale factors, the PE is calculated for the time evolution and the obtained values are shown in figure 5. It is observed, from Figure 5, that, as the value of scale factor is increased, the value of PE reduces from 1 to 0, which denotes that the intrinsic chaos of the oscillators is getting suppressed and the system has begun to oscillate in the relaxation oscillation regime.

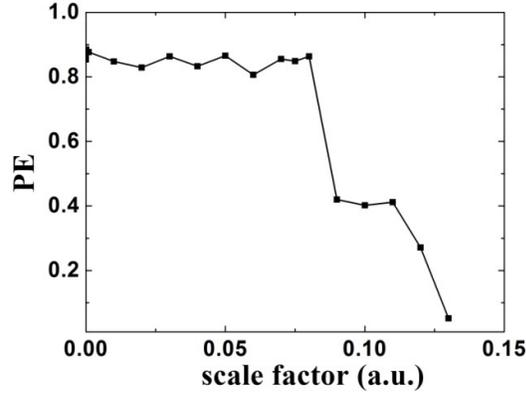

**Figure 5**: Permutation Entropy (PE) calculated at different scale factor (sf) values.

In the evolution of error variables, we showed using the Lyapunov stability analysis, the point $(E_1, E_2, E_3) = (0,0,0)$ is stable. This stability is necessary for achieving synchronization between the two systems. In the next section, we present our numerical findings on the character of synchronization of state variables under parametric control.

## 4. Co-existence of Synchronization and anti-synchronization

The dynamics of all the three state variables are considered for this study. We vary the parameter $\sigma$ of the control functions between 0 and 1 and seek for possible states of synchronization. In Figure 6 (a-f), we present the temporal evolution of error dynamics (defined by equation 4) along with the temporal evolution of state variables in the inset. The corresponding synchronization plots are shown in figure 6(g-l). It can be seen that all the three error variables asymptotically goes to zero. This is in agreement with the result of Lyapunov stability analysis of section 2, where we showed that $(E_1,E_2,E_3)=(0,0,0)$ is stable.

For $\sigma = 0.0$, the error function evolution for the first, second and third state variables are shown in figure 6(a), 6(b) and 6(c) respectively. We have kept the scale factor to be 0.07 for this study. The corresponding inset shows the temporal evolution of the state variables.

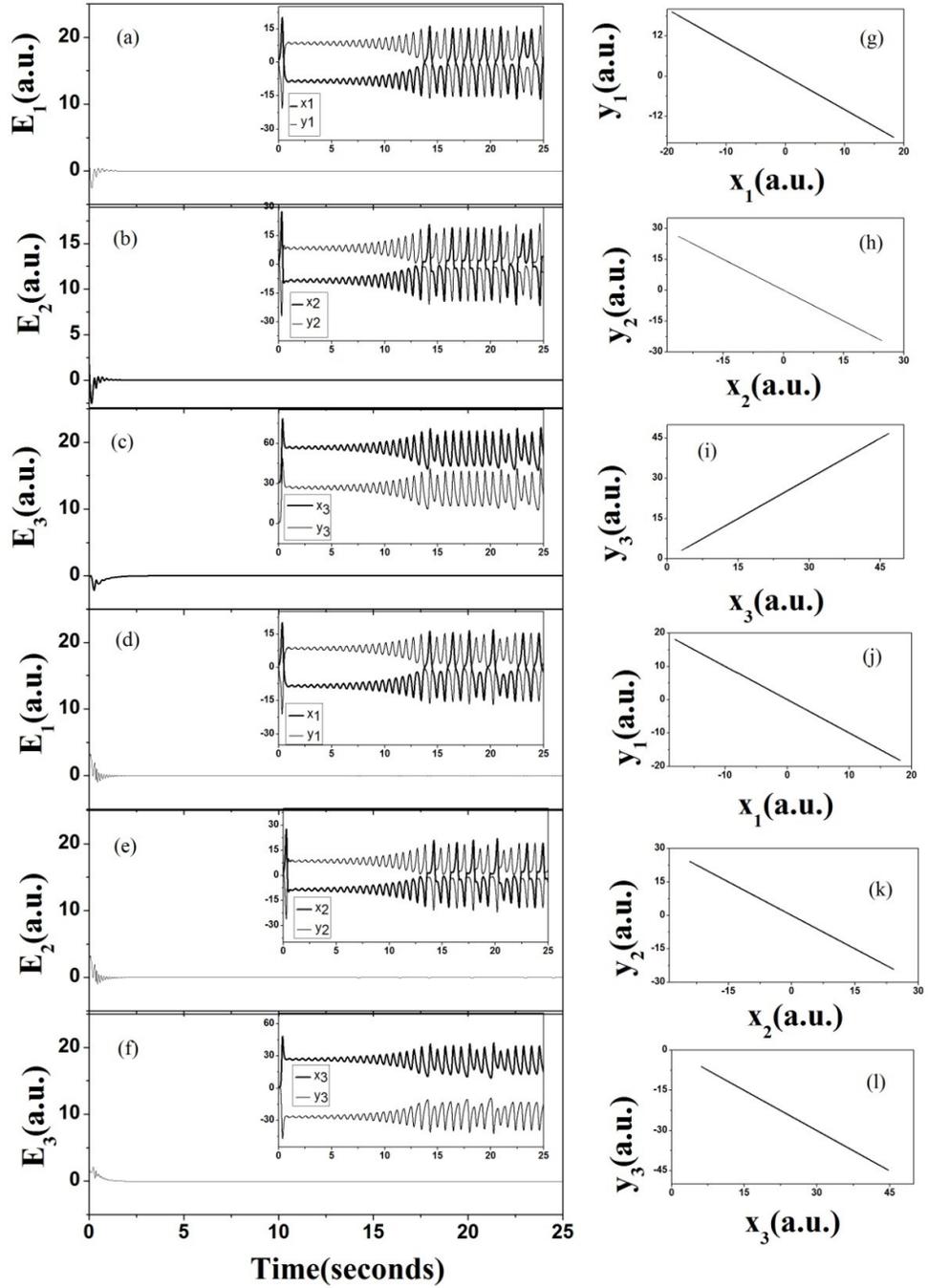

**Figure 6 (a)-(c)** are the error dynamics of the GLS for σ=0.0, **(d)-(f)** corresponds to σ=1.0 (the insets are the time evolution of corresponding state variables),**(g)–(i)** are the synchronization plots for σ =0.0 ,**(j)-(l)** corresponds to σ =1.0

The first ($x_1$ , $y_1$) and second ($x_2$ and $y_2$) state variables exhibit anti-synchronization as shown in figures 6(g) and 6(h). Figure 6(i) infers the synchronization between the state variables $y_3$ and $x_3$. On varying the value of the parameter σ from 0 to 1, the nature of the dynamics of the third state variable, $x_3$ and $y_3$ changes such that they exhibit anti-synchronization [33-37], as is evident from figure 6(l).

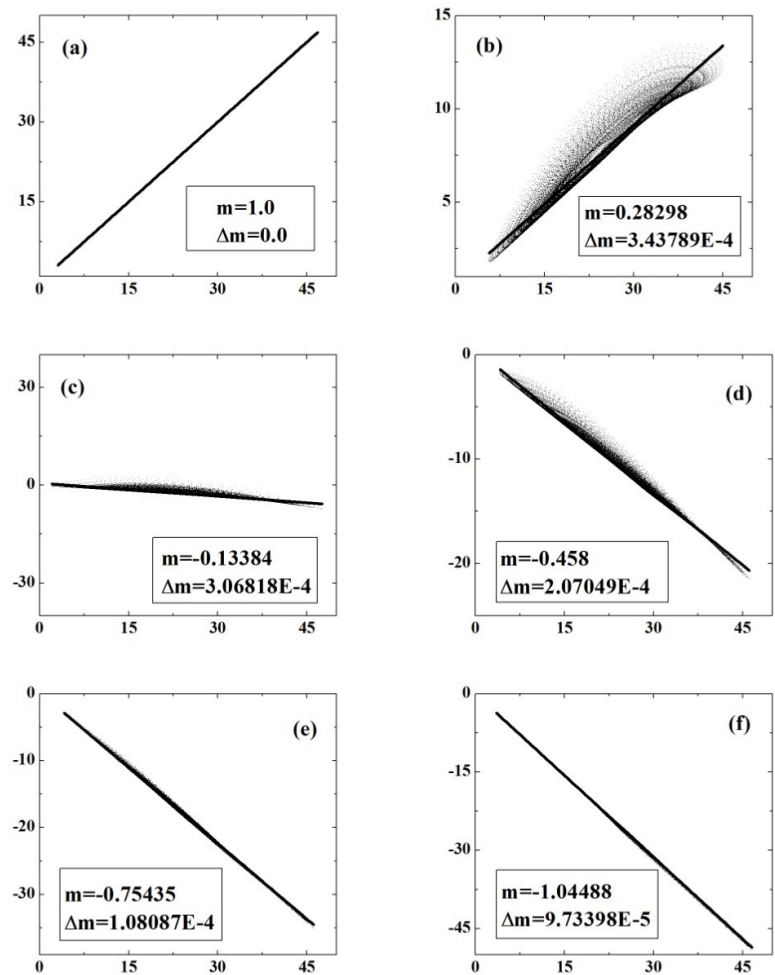

**Figure 7**: Synchronization plots between $x_3$ and $y_3$, σ is varied from 0.0 to 1.0, **(a)** σ = 0.0, **(b)** σ = 0.2, **(c)** σ = 0.4, **(d)** σ = 0.6, **(e)** σ = 0.8, **(f)** σ = 1.0

In figure 7, the synchronization plots between $y_3$ and $x_3$ are plotted for various values of control parameter σ. As σ is varied from 0 to 1 in steps of 0.2, as seen from figures 7(a) to 7(f), the slope of synchronization plot shows a transition from having a positive slope to negative slope implying a transition from synchronization to anti-synchronization.

The synchronization plots are fitted to a straight line and the magnitude of the slope (m) is obtained. When σ = 0, m is found to be unity and it reduced to 0.28 when σ is 0.2. For σ = 0.4, the value of slope acquires a negative value and m = -0.134, implying that state variables are getting into the regime of anti-synchronization.

On further increasing the value of σ, the slope value tends to become more negative and at σ = 1.0 the synchronization plot is having a negative slope of magnitude m = -1.045. This shows the continuous change-over of synchronization to anti-synchronization of state variables, enabled by varying a control parameter (σ).

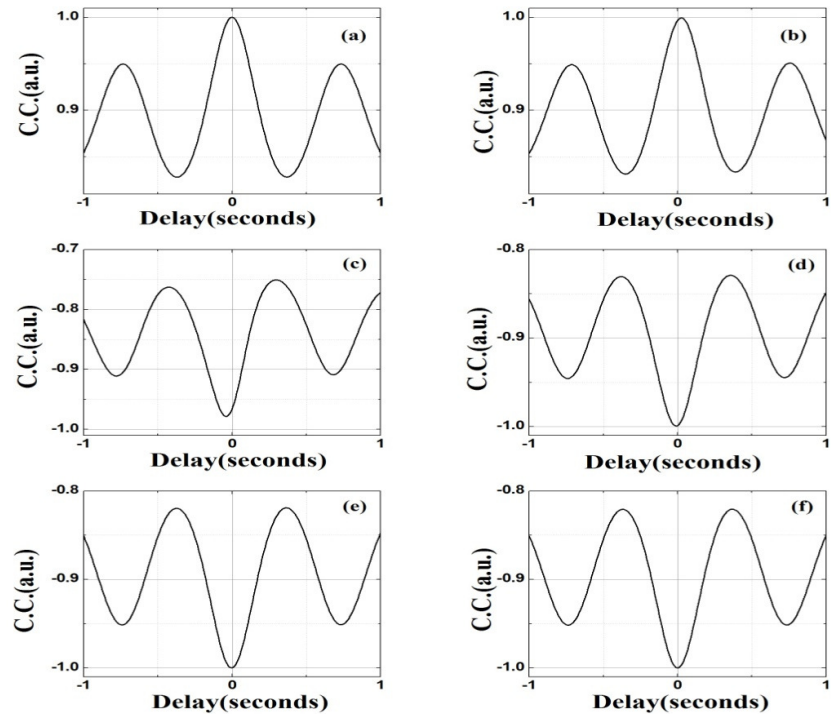

**Figure 8**: Cross correltaion plots plotted between $x_3$ and $y_3$, σ is varied from 0.0 to 1.0, **(a)** σ=0.0, **(b)** σ=0.2, **(c)** σ=0.4, **(d)** σ=0.6, **(e)** σ=0.8, **(f)** σ=1.0.

Cross-correlation analysis between the state variables $x_3$ and $y_3$ has been carried out for different values of sigma ($\sigma$) and the results are shows in figure 8. From the plots we observe that for values of $\sigma$, resulting in positive slope synchronization, the maximum of cross-correlation is at zero delay. But cross-correlation is a minimum for the cases corresponding to negative sloped synchronization plots as shown in Figure 8. Thus the cross-correlation plots affirm the transition of the nature of synchronization from synchronization to anti-synchronization.

## 5. CONCLUSION

Two Lorenz oscillators are coupled mutually via non-linear control functions. The temporal evolution of the state variables is studied and their synchronization properties are investigated. The strength of coupling, as controlled by a scale factor, is varied (increased) which is found to result in a reduction in the complexity of chaotic dynamics while preserving synchronization. Thus we are able to suppress chaos which preserving synchronization between mutually coupled Lorenz oscillators. The chaos suppression is validated by obtaining the power spectra, Lyapunov exponents and permutation entropy. We also show the possibility of transition from synchronization to anti-synchronization of a state variable by way of parametric control.

**ACKNOWLEDGEMENTS:**

Authors thank Dr. E. Jayaprasath for the help in permutation entropy calculations.